\documentclass[%
 superscriptaddress,
 preprint,
 showpacs,preprintnumbers,
 amsmath,amssymb,
 aps,
 prb,
]{revtex4-1}

\usepackage{graphicx}
\usepackage{dcolumn}
\usepackage{bm}

\usepackage{upgreek}
\usepackage{amsmath}

\begin{document}


\title{Enhancing Goos-Hänchen shift based on magnetic dipole quasi-bound states in the continuum in all-dielectric metasurfaces}

\author{Zhiwei Zheng}
\affiliation{Institute for Advanced Study, Nanchang University, Nanchang 330031, China}
\affiliation{Jiangxi Key Laboratory for Microscale Interdisciplinary Study, Nanchang University, Nanchang 330031, China}

\author{Ying Zhu}
\affiliation{Institute for Advanced Study, Nanchang University, Nanchang 330031, China}
\affiliation{Jiangxi Key Laboratory for Microscale Interdisciplinary Study, Nanchang University, Nanchang 330031, China}

\author{Junyi Duan}
\affiliation{Institute for Advanced Study, Nanchang University, Nanchang 330031, China}
\affiliation{Jiangxi Key Laboratory for Microscale Interdisciplinary Study, Nanchang University, Nanchang 330031, China}

\author{Meibao Qin}
\affiliation{Department of Physics, Nanchang University, Nanchang 330031, China}

\author{Feng Wu}
\email{fengwu@gpnu.edu.cn}
\affiliation{School of Optoelectronic Engineering, Guangdong Polytechnic Normal University, Guangzhou 510665, China}

\author{Shuyuan Xiao}
\email{syxiao@ncu.edu.cn}
\affiliation{Institute for Advanced Study, Nanchang University, Nanchang 330031, China}
\affiliation{Jiangxi Key Laboratory for Microscale Interdisciplinary Study, Nanchang University, Nanchang 330031, China}

\begin{abstract}
Metasurface-mediated bound states in the continuum (BIC) provides a versatile platform for light manipulation at subwavelength dimension with diverging radiative quality factor and extreme optical localization. In this work, we employ magnetic dipole quasi-BIC resonance in asymmetric silicon nanobar metasurfaces to realize giant Goos-Hänchen (GH) shift enhancement by more than three orders of wavelength. In sharp contrast to GH shift based on the Brewster dip or transmission-type resonance, the maximum GH shift here is located at the reflection peak with unity reflectance, which can be conveniently detected in the experiment. By adjusting the asymmetric parameter of metasurfaces, the $Q$-factor and GH shift can be modulated accordingly. More interestingly, it is found that GH shift exhibits an inverse quadratic dependence on the asymmetric parameter. Furthermore, we design an ultrasensitive environmental refractive index sensor based on the quasi-BIC enhanced GH shift, with a maximum sensitivity of 1.5$\times$10$^{7}$ $\upmu$m/RIU. Our work not only reveals the essential role of BIC in engineering the basic optical phenomena, but also suggests the way for pushing the performance limits of optical communication devices, information storage, wavelength division de/multiplexers, and ultrasensitive sensors.
\end{abstract}

\maketitle


\section{\label{sec1}Introduction}
Metasurfaces are two-dimensional artificial periodic arrays of subwavelength optical resonators, which can realize flexible and effective regulation of electromagnetic wave amplitude, phase, and polarization\cite{Liu2011, Zheludev2012, Xiao2020}. In recent years, metasurfaces made of dielectric materials have emerged as a competitive rival to their metal counterparts since they can avoid the annoying non-radiative loss and show a high diversity of available functionalities with near-unity efficiency. High $Q$-factor guided mode\cite{Bezus2018, Maksimov2020, Cao2020}, trapped mode\cite{Zhang2013, Sayanskiy2019, Xu2019, Zhou2020}, toroidal mode\cite{Tuz2018, He2018, Zhou2019}, and supercavity mode\cite{Kodigala2017, Rybin2017, Kyaw2020, Huang2021} can be excited in dielectric metasurfaces, manifesting as Fano resonance which essentially links to the physical concept of bound state in continuum (BIC)\cite{Hsu2016, Koshelev2018, Sadreev2021}. The sharp spectral feature and giant local field enhancement has a wide application prospect in optical devices such as light absorbers\cite{Wang2019, Tian2020, Wang2020}, emitters\cite{Zhu2020, Muhammad2021}, coupled optical waveguides\cite{Bezus2018a, Bykov2019}, nonlinear harmonic generation\cite{Carletti2018, Volkovskaya2020, Anthur2020, Ning2021}, and sensors\cite{Tittl2018, Romano2018, Wu2019, Ndao2020}.

The Goos-Hänchen (GH) shift was first experimentally observed by F. Goos and H. Hänchen in 1947\cite{Goos1947}. When a beam of light is totally reflected at the interface between two kinds of media, the reflected light will have a tiny lateral shift in the incidence plane. Since the discovery of the GH shift, it has attracted a great deal of attention because of its profound physical significance and broad application scenarios\cite{Wang2013, Wu2020}. Without any assistance of external enhancements, the GH shift is only a few times the length of its wavelength. There are several physical mechanisms that can be resorted to for enhancing the GH shift. The first is based on the Brewster effect. In the vicinity of the Brewster angle, the reflection phase will vary dramatically with the incident angle, which will results in a large GH shift\cite{Lai2002, Huang2011}. The second is based on the transmission-type resonances. A high $Q$-factor resonance also leads to dramatic change in the reflection phase around the resonant peak\cite{Kaiser1996}. Based on these, numerous theoretical and experimental works have been devoted to enhancing the GH shift on different structure interfaces such as weakly absorbing slab\cite{Wang2005, Shen2006, Wang2018}, surface plasmon resonator\cite{Yin2006, Yallapragada2016, You2019, Petrov2020}, Bloch surface wave resonator\cite{Wan2012, Kong2018, Wang2021}, dielectric metagrating and photonic crystal slab\cite{Wang2006, Wong2018}. However, there is a common defect in these two enhancement mechanisms: although the maximum GH shift can be enhanced to the length of orders of wavelength, it is exactly located at the reflection dip with extremely low reflectance, which is difficult to be detected. Therefore, enhancing the GH shift while simultaneously maintaining the unity reflectance remains elusive.

In this work, we introduce the metasurface-mediated BIC as a high-efficiency platform to realize the GH shift enhancement. The asymmetric silicon nanobar metasurfaces supports a reflection-type magnetic dipole quasi-BIC resonance with ultranarrow spectral width, which results in sharply varying angular deviations of the phase of the incident light and therefore a giant GH shift of three orders of wavelength. Since the enhancement of the GH shift is achieved at the total reflection configuration, the reflected beam can be easily detected in experiment. The $Q$-factors and GH shift can be modulated by adjusting the structural asymmetric parameter. In addition, we find that the GH shift exhibits an inverse quadratic dependence on the asymmetric parameter which follows the similar relationship between $Q$-factor and asymmetric parameter in BIC formalism. Finally, as an example of application, we design an ultrasensitive environmental refractive index sensor based on the quasi-BIC enhanced GH shift, and the maximum value of the sensitivity reaches as high as 1.5$\times$10$^{7}$ $\upmu$m/RIU.

\section{\label{sec2}Bound states in the continuum supported in all-dielectric metasurfaces}
The all-dielectric metasurfaces utilized for enhancing the GH shift are schematically plotted in Fig. 1(a). Silicon with the refractive index $n=3.5$ is chosen as the constituent material due to its negligible loss in the wavelength range of interest. The specific geometrical parameters are labeled in Fig. 1(b). The unit cell is arranged with a square period $P=900$ nm, and composed of a pair of asymmetric nanobars with a fixed interval of $D=250$ nm. The width and thickness of the nanobars are $W=200$ nm and $H=160$ nm, respectively. The lengths are set with a fixed $L_{1}=750$ nm and a variable $L_{2}$ for each case studied. The asymmetric parameters $\alpha$ is thus defined as $\Delta L/L_{1}$, where $\Delta L$ is the length difference between the nanobars. For different $L_{2}$, $\alpha$ can vary between 0 and 1.
\begin{figure}[h!]
	\centering\includegraphics[width=12cm]{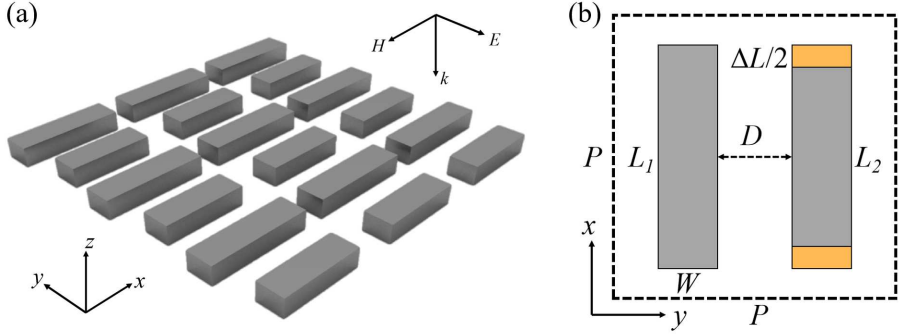}
	\caption{(a) The schematic of the proposed all-dielectric metasurfaces. (b) The unit cell is arranged with a period $P=900$ nm, and composed of a pair of asymmetric nanobars with a fixed interval of $D=250$ nm. The width and thickness of the nanobars are $W=200$ nm and $H=160$ nm. $L_{1}=750$ nm and $L_{2}$ represent the lengths of the long nanobar and short nanobar, respectively. $\Delta L$ is the length difference between the nanobars.}
\end{figure}

Numerical simulations are performed with the finite element method (FEM) via COMSOL Multiphysics. A TM-polarized plane wave (magnetic field parallel to the $x$ direction) is incident along the -$z$ axis. The Floquet periodic boundary
conditions are employed in $x$ and $y$ directions, and the perfectly matching layers are adopted in the $z$ direction. The incident angle $\theta$ is initially set to $5^{\circ}$ and the length of short nanobar $L_{2}$ is set to 600 nm ($\alpha= 0.2$). The genuine BIC transforms to the quasi-BIC by breaking such in-plane symmetry of the nanobar pair, which opens a coupling channel to the radiative continuum, and manifests itself in the reflection spectrum as a Fano resonance. As shown in Fig. 2(a), a pronounced resonance peak is located at the wavelength $\lambda_{0}=1049.76$ nm with an ultranarrow full width at half maximum (FWHM) of 0.13 nm, which is well fitted by the classical Fano formula within the coupled-mode theory\cite{Fan2003}. Furthermore, the far-field and near-field distributions are respectively calculated to identify the mode origin of the Fano resonance. The contributions of multipole moments to the far-field are decomposed at the resonance, as shown in Fig. 2(b). The dominant component is provided by the magnetic dipole (MD), which is at least twice as much as the electric quadrupole (EQ), the toroidal dipole (TD), and is stronger than the magnetic quadrupole (MQ) and electric dipole (ED) by orders of magnitude. Besides, the near-field distributions are in strict accordance with the far-field. As shown in Fig. 2(c), the currents in the two nanobars oscillate in opposite directions and generate two anti-parallel magnetic dipoles, giving rise to a dominant component of the magnetic dipole, and a smaller amount of the toroidal dipole and electric quadrupole viewed as a circular head-to-tail arrangement of magnetic dipoles.
\begin{figure}[h!]
	\centering\includegraphics[width=13cm]{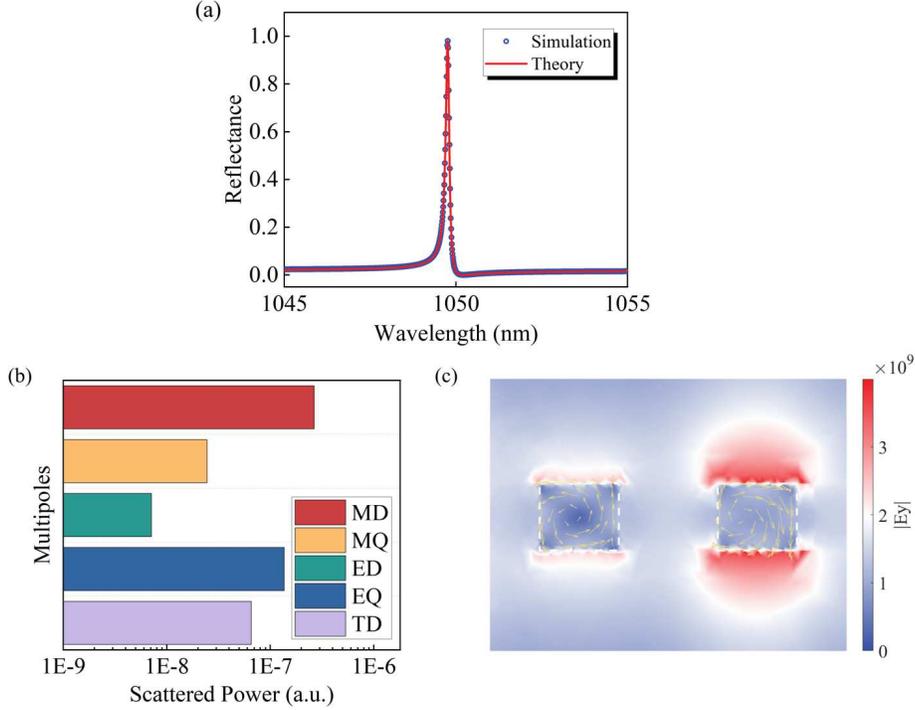}
	\caption{(a) The simulated and theoretically fitted reflection spectra when the length of short nanobar is set to $L_{2}=600$ nm and incident angle is set to $\theta=5^{\circ}$. (b) The contributions of multipole moments to the far-field at resonance. (c) The $y$-component of electric field corresponding to the reflection peak, overlaid with arrows indicating the direction of displacement current.}
\end{figure}

Such magnetic dipole quasi-BIC resonance is involved with the external radiative channel, and the $Q$-factor and the resonance spectral width can be flexibly modulated through adjusting the structural asymmetry parameter. In this case, we obtain a series of $Q$-factors through sweeping the length of the short nanobar $L_{2}$ to change the asymmetry parameter $\alpha$. The dependence of $Q$-factor on $\alpha$ is shown in Fig. 3(a). The insets show the exact structure corresponding to the cases $\alpha=0.1$ and 0.6, respectively. When $\alpha=0.1$, the $Q$-factor reaches as high as $2.91\times10^{4}$. In theory, the $Q$-factor can approximate to infinity as $\alpha$ infinitely approaches to zero which corresponds to the genuine BIC. Furthermore, we plot the $Q$-factor as a function of $\alpha^{-2}$, as shown in Fig. 3(b). The $Q$-factor is almost proportional to $\alpha^{-2}$, i.e., $Q\propto \alpha^{-2}$, which is consistent with the theoretical formalism for asymmetric resonances governed by BIC. It reveals the fact that the $Q$-factor has a huge range of linear variation by modulating the asymmetry parameter, providing an efficient strategy to design the structure for the GH shift enhancement.
\begin{figure}[h!]
	\centering\includegraphics[width=13cm]{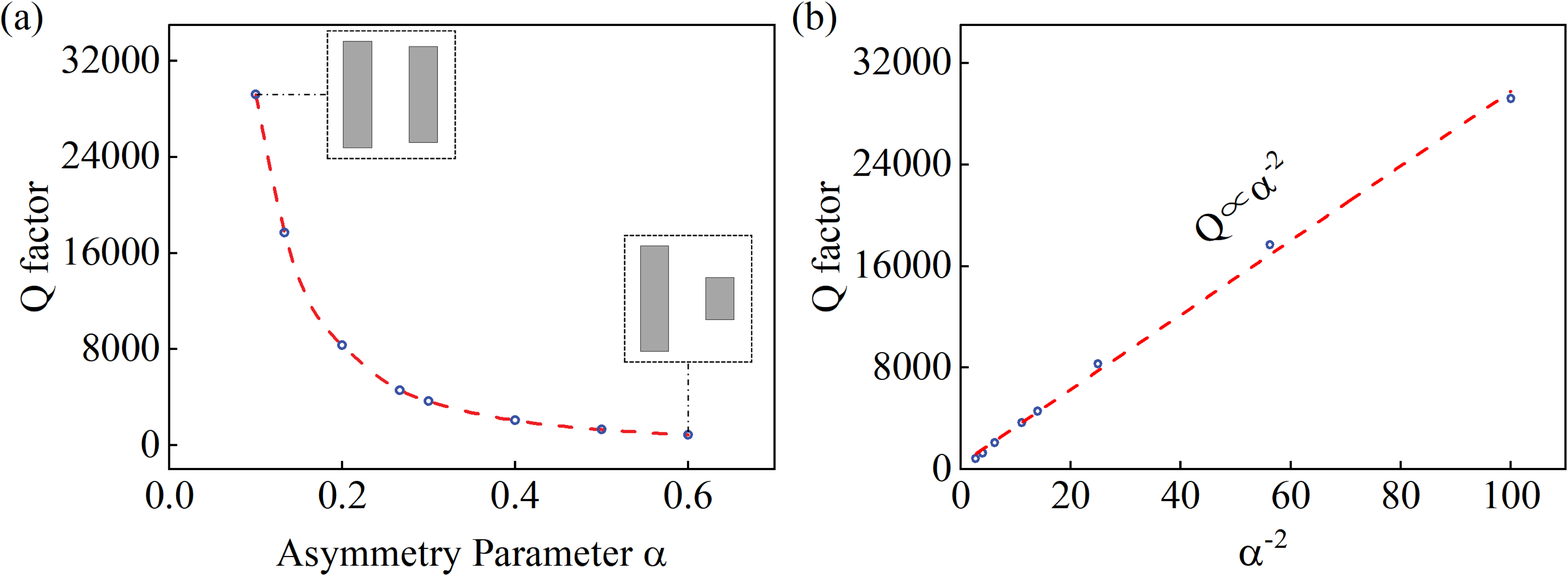}
	\caption{(a) The dependence of $Q$-factor on the asymmetry parameter $\alpha$. The insets show the specific structures with corresponding to $\alpha$. (b) The dependence of $Q$-factor on $\alpha^{-2}$ with the red dashed line representing linear fitting.}
\end{figure}

\section{\label{sec3}Enhancing GH shift based on quasi-BIC resonance}
In this section, the lengths of the short nanobar are set to $L_{2}=600$ nm, 525 nm, and 450 nm ($\alpha=0.2$, 0.3, and 0.4), respectively. Meanwhile, the corresponding incident wavelengths $\lambda_0$ are chosen at the resonance peak under oblique incidence $\theta=5^{\circ}$. By sweeping the incident angle from $3^{\circ}$ to $7^{\circ}$, we obtain the reflectance angular spectra which are shown with the green lines in Figs. 4(a), 4(c), and 4(e), respectively. The blue lines represents the corresponding reflection phase angular spectra. It can be seen that the reflection phase changes drastically around the incident angle $\theta=5^{\circ}$. According to the stationary phase method proposed by K. Artmann, the GH shift is proportional to the partial derivative of the reflection phase to the incident angle\cite{Artmann1948},
\begin{equation}
	S_{\text{GH}} =
	-\frac{\lambda_{0}}{2\pi} \frac{\partial\varphi(\theta)}{\partial\theta},
\end{equation}
where $\lambda_0$ is the resonance wavelength, $\varphi$($\theta$) represents the reflection phase. Through this formula it can be inferred that the gradient of the reflection phase directly determines the magnitude of the GH shift, and a high $Q$-factor resonance with incidence angle sensitivity will lead to GH shift enhancement to a great extent. The GH shift angular spectra are subsequently calculated and shown in Figs. 4(b), 4(d), and 4(f), respectively. Around the incident degree of the resonance peak $\theta=5^{\circ}$, the maximum values of the GH shift are obtained in each case. Because the $Q$-factor and the incidence angle sensitivity of the quasi-BIC resonance highly depends on the asymmetry parameter $\alpha$, the maximum GH shift shows an obvious positive relationship with the asymmetry parameter $\alpha$. When $\alpha$ starts at 0.4, the maximum GH shift is about $429\lambda_{0}$. As $\alpha$ decreases to 0.3, the maximum GH shift goes up to $761\lambda_{0}$. Finally when $\alpha$ comes to 0.2, the maximum GH shift reaches as high as $1654\lambda_{0}$, which is more than three orders of wavelength. In sharp contrast to that based on the Brewster angle or transmission-type resonance, the GH shift here is exactly at the reflection peak with the unity reflectance and can be easily detected in practical measurement. 
\begin{figure}[h!]
	\centering\includegraphics[width=14cm]{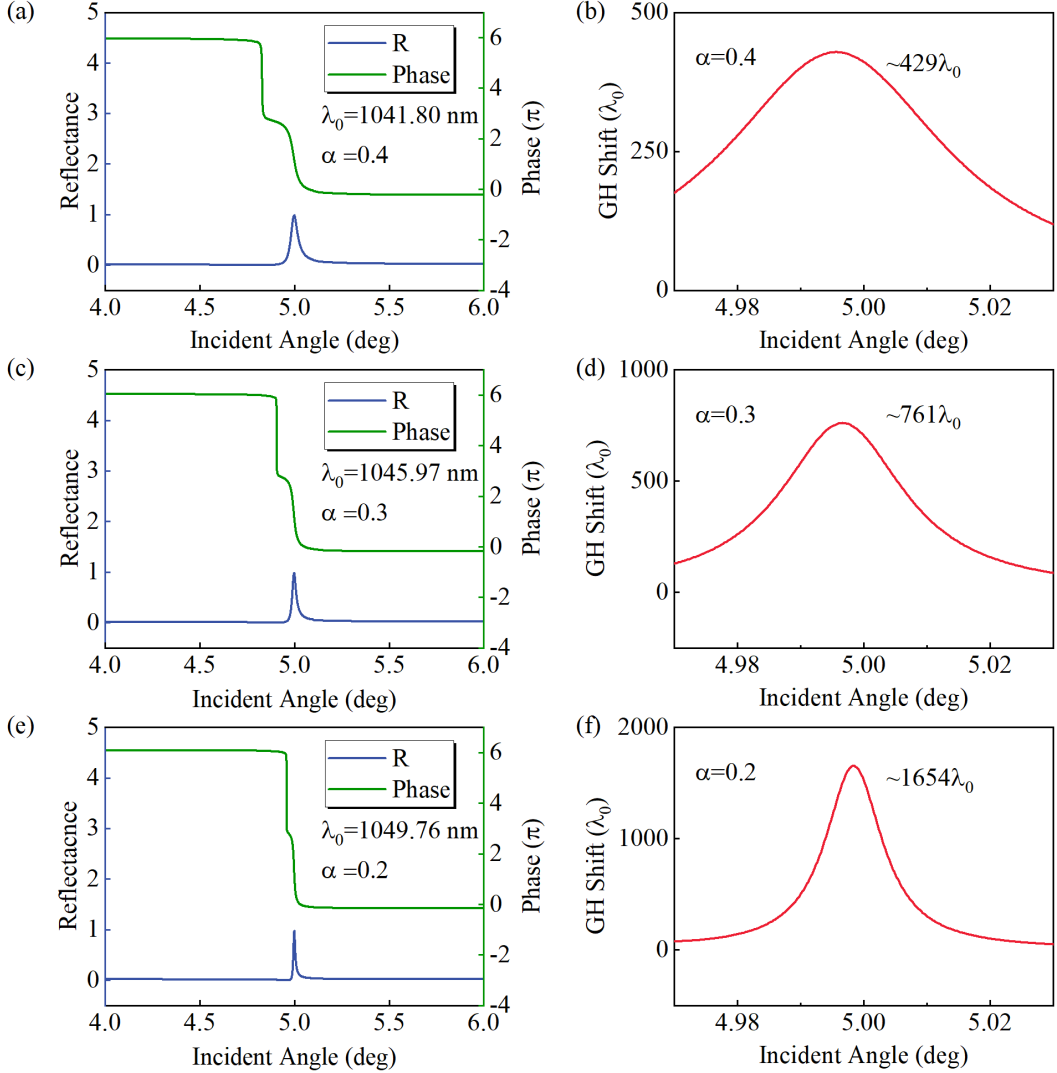}
	\caption{The reflectance angular spectra (blue line) and phase angular spectra (green line) for (a) $\alpha$ = 0.4 at $\lambda_0$ = 1041.80 nm, (c) $\alpha$ = 0.3 at $\lambda_0$ = 1045.97 nm, and (e) $\alpha$ = 0.2 at $\lambda_0$ = 1049.76 nm. The GH shift angular spectra around the incident angle $\theta=5^{\circ}$ for (b) $\alpha$ = 0.4, (d) $\alpha$ = 0.3, and (f) $\alpha$ = 0.2.}
\end{figure}

In addition, we further provide the dependence of the maximum value of the GH shift at the resonance on the the asymmetry parameter as shown in Fig. 5. It is interesting to notice that the GH shift is almost proportional to $\alpha^{-2}$, following the similar relationship between $Q$-factor and asymmetric parameter in BIC theory, which offers a concise way to directly control the GH shift enhancement by simply adjusting the asymmetric parameter of the all-dielectric metasurfaces.
\begin{figure}[h!]
	\centering\includegraphics[width=6.5cm]{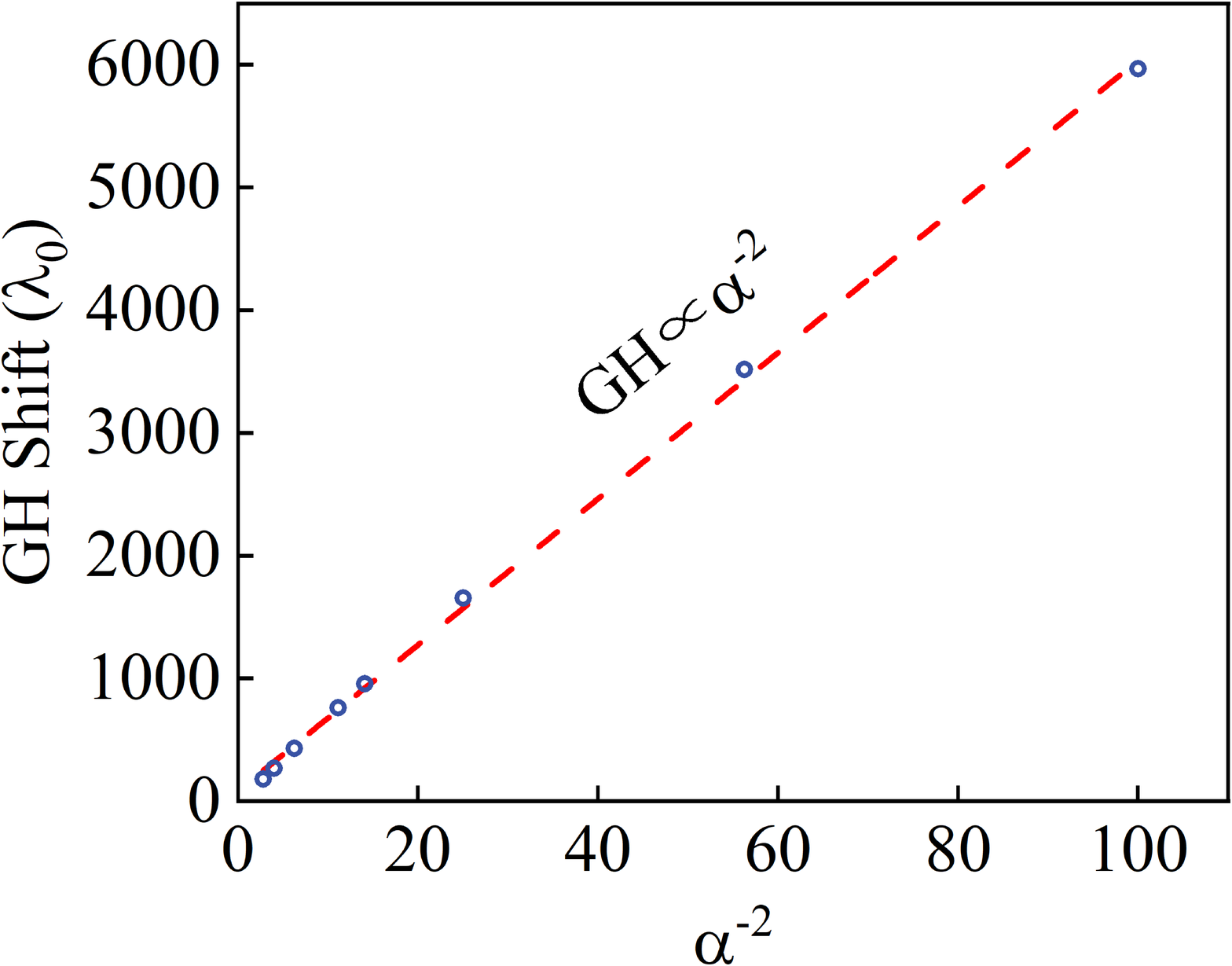}
	\caption{The dependence of the maximum value of the GH shift on the the asymmetry parameter $\alpha^{-2}$.}
\end{figure}

\section{\label{sec4}Refractive index sensor based on giant GH shift}
Due to the ultrahigh sensitivity of the GH shift, we apply it to the refractive index sensing to detect any tiny change in the environmental refractive index. In many application scenarios such as climate monitoring and biochemical detecting, the refractive index sensors are highly demanded\cite{Wang2008}.

Here, the length of the short nanobar is fixed to $L_{2}=600$ nm maintaining the resonance wavelength $\lambda_{0}=1049.76$ nm. Fig. 6(a) gives the GH shift angular spectrum at environmental refractive index $n_{\text{e}}=1$ and 1.00005, as shown by the blue line and red line, respectively. As $n_{\text{e}}$ increases from 1 to 1.00005, the spectral line shifts towards the smaller incident angle. Around the incident angle of the resonance peak $\theta=5^{\circ}$, the GH shift changes from the maximum value 1736.6 $\upmu$m to 1627.2 $\upmu$m. The giant variation of the GH shift 109.4 $\upmu$m with the tiny change in the refractive index 0.00005 provides a basis for the application of refractive index sensing. The blue line in Fig. 6(b) shows the GH shift as a function of the environmental refractive index $n_{\text{e}}$. When $n_{\text{e}}$ varies between 1.00001 and 1.0002, the GH shift increases to a maximum value 1736.5 $\upmu$m and then monotonically decreases to 284.4 $\upmu$m. 
\begin{figure}[h!]
	\centering\includegraphics[width=14cm]{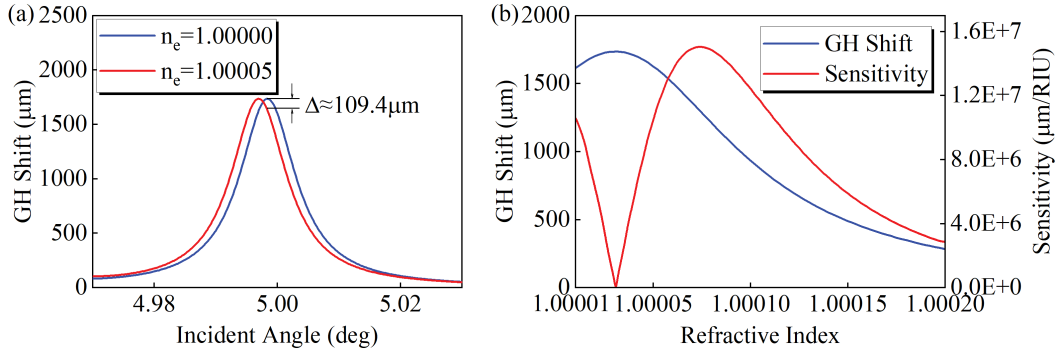}
	\caption{(a) The GH shift angular spectrum at environmental refractive index $n_{\text{e}}=1$ (blue line) and 1.00005 (red line). (b) The GH shift and sensitivity of the proposed sensor as a function of the environmental refractive index $n_{\text{e}}$.}
\end{figure}

It can be intuitively observed the fluctuation of the GH shift caused by the change in environmental refractive index, and thus enables us to directly find the corresponding refractive index by measuring the GH shift. In addition, we define the sensitivity of the refractive index sensor we proposed as the derivative of the GH shift (absolute value) with respect to the refractive index, as shown by the red line in Fig. 6(b). The maximum value of the sensitivity reaches 1.5$\times10^{7}$ $\upmu$m/RIU around $n_{\text{e}}=1.000075$, which indicates that it is highly efficient and sensitive to the change in the environmental refractive index in a certain range. In addition, by simply changing the wavelength of the incident light, the high sensitivity can be obtained in other specific ranges as well.

\section{\label{sec5}Conclusions}
In conclusions, we realize giant enhancement of the GH shift based on reflection-type quasi-BIC in a typical all-dielectric metasurfaces which supports magnetic dipole resonance. Due to the ultrahigh Q-factor and the incidence angle sensitivity, the GH shift can reach three orders of wavelength, which can be easily detected. In the meanwhile, we also bridge a linear link between the GH shift and geometric asymmetric parameter of the metasurfaces. Such behavior follows the similar dependence of $Q$-factor on $\alpha^{-2}$ in the BIC theoretical formalism. Finally, the structure is applied to environmental refractive index sensing with a maximum sensitivity of 1.5$\times$10$^{7}$ $\upmu$m/RIU. This work paves an efficient and smart way towards the design of optical switches, light information storage devices, and some kinds of highly sensitive sensors including refractive index sensors, temperature sensors, and solution concentration sensors.

\begin{acknowledgments}	
This work is supported by the National Natural Science Foundation of China (Grant No. 11947065), the Natural Science Foundation of Jiangxi Province (Grant No. 20202BAB211007), the Interdisciplinary Innovation Fund of Nanchang University (Grant No. 2019-9166-27060003), and the Start-Up Funding of Guangdong Polytechnic Normal University (Grant No. 2021SDKYA033). 
\end{acknowledgments}

%

\end{document}